\documentclass[aps,twocolumn,superscriptaddress,pra,floatfix,showpacs]{revtex4}
\usepackage[usenames]{color}
\usepackage{amsmath}
\usepackage{amssymb}
\usepackage{mathrsfs}
\usepackage{euscript}
\usepackage{graphicx}
\usepackage{graphics}

\begin{document}
\title{\textit{ab initio} frame transformation calculations of direct and indirect dissociative recombination rates of HeH$^+$ + e$^-$}
\author{Daniel J. Haxton}
\email{dhaxton@jila.colodado.edu}
\affiliation{Department of Physics and JILA, University of Colorado, Boulder Colorado 80309}
\author{Chris H. Greene}
\email{chris.greene@colorado.edu}
\affiliation{Department of Physics and JILA, University of Colorado, Boulder Colorado 80309}

\pacs{03.65.Nk, 34.80.-i, 34.80.Lx, 33.20.Wr}

\begin{abstract}

The HeH$^+$ cation undergoes dissociative recombination with a free electron
to produce neutral He and H fragments.  We present calculations using 
\textit{ab initio} quantum defects and Fano's rovibrational frame transformation
technique, along with the methodology of Ref.~\cite{hamilton}, 
to obtain the recombination rate both in the low-energy (1-300~meV) and
high-energy (ca. 0.6 hartree) regions.  We obtain very good agreement with
experimental results, demonstrating that this relatively simple method is
able to reproduce observed rates for both indirect dissociative recombination,
driven by rovibrationally autoionizing states in the low-energy region, 
and direct dissociative recombination, driven by electronically autoionizing
Rydberg states attached to higher-energy excited cation channels.

\end{abstract}

\maketitle

\section{Introduction}

The dissociative recombination (DR)\cite{guberman,mitchell,larssonreview}
 reaction HeH$^+$ + e$^-$ $\rightarrow$ He + H 
has received considerable theoretical and experimental interest.
HeH$^+$ ion chemistry is important to the
understanding of the composition of interstellar space.  
Observations~\cite{moorhead} have failed to show much HeH$^+$ present in the
interstellar medium, and were in contradiction to the early 
prediction~\cite{roberge} that this species would be abundant.
These observations indicated that the low-energy DR cross section
of HeH$^+$ + e$^-$ is indeed significant, though there is no valence
electronic state of the neutral that provides a mechanism for the high rate
at low collision energy.

Thus, much theoretical and experimental interest has been focused
on this process in the intervening years.
Experiments \cite{yousif, tanabe0,sundstrom, 
tanabe0b,mowat,semaniak,stromholm, tanabe} 
show peaks in the DR cross section 
in both the low-energy region (ca. 1-300~meV) and the high-energy region around
20~eV.  It is the low-energy peak that is responsible for the destruction
of HeH$^+$ in interstellar space.

\begin{figure}[t]
\resizebox{0.9\columnwidth}{!}{\includegraphics*[0.2in,0.6in][5.7in,4.2in]{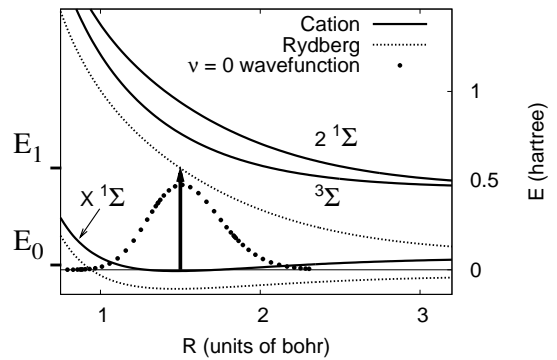}}
\caption{Schematic of HeH$^+$ DR.  The ground rovibrational state
of the cation is plotted
as dots; the cation curves, as solid lines; and Rydberg curves 
supported by the lowest and highest cation curves, as dotted lines.  Typical
incident electron energies for the low-energy and high-energy DR 
are labeled E$_0$ and E$_1$, respectively.
\label{schematic}}
\end{figure}

The low and high energy peaks are caused by two different mechanisms, 
the indirect and direct, respectively.  The indirect mechanism has
proved more difficult to treat theoretically, and the initial 
underestimate~\cite{roberge} of the low-energy DR rate was based 
upon a consideration of the direct mechanism only.  

The presence of the
two mechanisms, and the importance of this process to the understanding
of interstellar chemistry, makes the DR of HeH$^+$ a prime target for
theory.  Several treatments~\cite{guberman,rescignoheh,larsonorel, wavepacket,takagi2004, takagi2005, tanabe} have been presented, and these have had
varying success at reproducing experimental DR rates.  The study of
Orel \textit{et al.}~\cite{rescignoheh} reproduced the magnitude and
size of the high-energy peak, but did not accurately reproduce the shoulder
on the high-energy side.  Takagi~\cite{takagi2004,takagi2005} has obtained
good agreement for the low-energy peak, but did not reproduce all the 
peaks and dips in the cross section perfectly.

As schematic of the process is shown in Fig.~\ref{schematic}.
Indirect DR\cite{guberman} in the low-energy region is driven by
transitions to 
rovibrationally autoionizing Rydberg states supported by the lowest
cation curve.  Such states are described as 
a Rydberg electron attached to
a rotationally or vibrationally excited, but electronically ground-state, 
HeH$^+$ cation core.   The DR process is driven by nonadiabatic coupling
to and among these Rydberg states.  The DR cross
section exhibits peaks corresponding to the locations of the autoionizing
resonances, which serve as doorways.

The direct DR process\cite{bates,omalley} 
is driven by a transfer of energy among the electronic
degrees of freedom only: an incident electron with sufficient energy may
excite the ground-state ( 1~$\Sigma$ ) HeH$^+$ cation core, such that the incident
electron becomes
trapped in a Rydberg state supported by a dissociative $\sigma$ cation
core.  This is a vertical transition, labeled by an arrow in Fig.~\ref{schematic},
 and nonadiabatic coupling is unimportant
to the total DR rate.  Once the electronic transition to the 
metastable Rydberg state has occurred, the molecule follows
the dissociative Rydberg potential energy surface toward breakup.
Our studies indicate that most of the high-energy DR to HeH$^+$ is 
accounted for by the Rydberg series converging to the 
upper singlet curve.

\section{Theory}

Several theoretical methodologies have found reasonable success in treating 
the dissociative recombination of HeH$^+$, but there is still room
for improvement.
Treatments\cite{rescignoheh,larsonorel, wavepacket} using the formalism 
of O'Malley\cite{omalley}, in which the DR rate is obtained 
by calculating potential energy
curves of the autoionizing Rydberg states of HeH, may be contrasted
with treatments\cite{takagi2004, takagi2005, tanabe} 
using multichannel quantum defect 
theory (MQDT)\cite{seaton,fanoft,changfano,general,general2} 
in which the only
potential energy curves explicitly included are those of the cation,
or those in which both cation and neutral curves are 
used~\cite{gusti,guberman}.
Scattering calculations using vibrational close coupling were used in
Ref.~\cite{sarpal}.  

The O'Malley treatment was used by Orel \textit{et al.}\cite{rescignoheh} 
to calculate
the high energy peak; those authors reproduced the magnitude and basic
shape of the high-energy peak obtained from the 1993 experiment 
of Sundstrom \textit{et al.}\cite{sundstrom} by
including a total of eight doubly excited Rydberg states (six $\Sigma$ and
two $\Pi$).
To calculate the low-energy peak under the O'Malley framework, 
Larson~\textit{et al.}\cite{larsonorel} 
required nonadiabatic
coupling matrix elements between the ground and $\Sigma$-core 
Rydberg states of HeH$^+$; their study underestimated the magnitude of the 
experimental result by a factor of ten to 100.

Treatments using multichannel quantum defect theory can be considerably
simpler than those using the O'Malley treatment because they do not 
necessarily require the 
enumeration of the Rydberg states contributing to the process.
Such treatments include ours and those of Takagi~\cite{takagi2004,takagi2005}
and Guberman~\cite{guberman}.  The main difference between our method and
those of Takagi and Gubmerman is that ours includes the DR process in the closed
channel space of the MQDT calculation, whereas the others include DR
channels in the open channel space.

Guberman's pioneering 1994 paper~\cite{guberman} 
employed the hybrid technique~\cite{gusti} incorporating 
both dissociative neutral and cation curves.  In this method one typically has dissociative valence states and MQDT S-matrices corresponding to the Ryberg series, and one proceeds by coupling these in the first Born approximation.  The coupling can be purely electronic or can arise from nonadiabatic coupling.  

Nonadiabatic coupling was employed in Guberman's treatment.  He defined MQDT S-matrices using the curve of the D state of neutral HeH.  He treated the C state, which lies below the D state, as an open dissociative channel.  The C and D states avoid and the nonadiabatic coupling between them is included as the discrete-continuum coupling as in Ref.~\cite{gusti}.

Guberman's technique has several advantages, notably that it incorporates the correct asymptotic form of the wavefunction.  However, it requires the identification of the dominant final electronic state channel of the DR process before doing the calculation, and thus may be difficult to apply to a general problem.

Our method of calculating DR rates using MQDT is similar to Takagi's
method\cite{takagi2004,takagi2005} in that it incorporates a vibrational
frame transformation\cite{fanoft,changfano} including both bound and
continuum vibrational wavefunctions of the HeH$^+$ cation.  Our methods
differ in the choice and treatment of the continuum vibrational states.
Takagi employs standing-wave boundary conditions (box states), and chooses
a subset of these states to represent the DR channels, interspersed among
others that are chosen to be closed.  We employ outgoing-wave boundary
conditions\cite{hamilton}, and consider such continuum channels open or 
closed according to the real part of their complex-valued energies.
Thus, in our method, the DR channels are not explicitly included in the
calculated s-matrix, whereas in Takagi and Gubmerman's treatments, they are.

Whereas Guberman's calculation incorporates the correct asymptotic form 
of the wavefunction, but is difficult to implement for a general system, 
our method and that of Takagi incorporate unphysical boundary conditions 
and are more easy to implement.  In Takagi's treatment, some box continuum 
states are chosen to be open and some closed, and these are interspersed 
among each other.  Therefore, in some box states the bound-state boundary 
condition is imposed for the electron, and for others, they are not.  The 
calculation has Rydberg states attached to closed box states and this 
situation is clearly unphysical.  In our calculation, we have wavefunctions 
with complex energy, asymptotically increasing at large bond length, which 
is also clearly unphysical.  However, in the limit that the discretized continuum 
states have a negligibly small imaginary component to their energy, our 
wavefunction corresponds to the correct physical scattering wavefunction 
having outgoing wave boundary conditions in the diatomic coordinate.

At present,
our method has not been applied to the problem of the final-state 
distribution of the electronic states of the fragment atoms.  Applying the
methods of Guberman or Takagi to this problem is more straightforward than it is for our method.

\subsection{Outgoing wave basis functions}

The original outline of our method\cite{hamilton} as well as 
subsequent calculations of DR rates for physical 
systems\cite{kokoo,roman,roman2} 
used the technique of Siegert pseudostates\cite{pseudo} to define the
outgoing-wave cation vibrational basis.  However, there are a few
difficulties with the formal theory presented in these papers
that we would like to address.  The Siegert-state basis is orthonormal
with respect to integration plus a surface term,
\begin{equation}
\int \ dR \ \chi_\alpha(R) \chi_\beta(R) \ + i \frac{ \chi_\alpha(R_0) \chi_\beta(R_0) }{ k_\alpha + k_\beta } \ = \ \delta_{\alpha \beta} \ ,
\end{equation}
where $k_\alpha$ is the Siegert pseudostate wavenumber eigenvalue
for state number $\alpha$.
By analogy, the frame-transformed S-matrix has been written (see, for
example, Eq.(6) in Ref.~\cite{hamilton})
\begin{equation}
\begin{split}
S_{\alpha \beta}  = & \int \ dR \ \chi_\alpha(R) s(R) \chi_\beta(R) \\
& + i \frac{ \chi_\alpha(R_0) s(R_0) \chi_\beta(R_0) }{ k_\alpha + k_\beta } \ .
\end{split}
\end{equation}
However, questions may be raised as to whether this  \textit{ad hoc} 
expression represents the 
proper transformation of the S-matrix from the $R$ basis to the $\chi$ basis.
For instance, this equation does not preserve the eigenphases of the MQDT s-matrix, and
is not invertible.  

The orthonormality properties of the Siegert pseudostates
are well known\cite{pseudo}, and it is not hard to derive a more 
appropriate expression for the transformed S-matrix using Siegert
pseudostates.  However, in the present paper we choose to use a
different method to enforce outgoing-wave boundary conditions in the
vibrational basis.

Complex absorbing 
potentials (CAP)~\cite{CAPform,CAPref1,CAPref2} or 
exterior complex scaling (ECS)\cite{abc1, abc2, moi2, moi3, moi4, moirev, ecs}
are alternative methods of defining outgoing-wave states.
The set of eigenvectors
obtained from these methods obey a simpler orthonormality relationship 
than do the Siegert pseudostates: they
may be chosen orthonormal with respect
to the inner product in which neither the bra nor ket is complex-conjugated,
called the C-norm,
\begin{equation}
\int \ dR \ \chi_\alpha(R) \chi_\beta(R) \ = \ \delta_{\alpha\beta} \ .
\end{equation}
Unlike the orthonormality relationship of the Siegert states, the C-norm
is an inner product and defines the ECS or CAP states as basis vectors
of a vector space.  The frame transformation of the fixed-nuclei S-matrix to
the ECS or CAP basis uses this inner product and is straightforward:
\begin{equation}
S_{\alpha \beta} = \int \ dR \ \chi_\alpha(R) s(R) \chi_\beta(R) \ .
\label{trans}
\end{equation}
This transformation 
preserves the eigenphases exactly in the complete-basis limit.  We then
apply the rotational frame transformation of Ref.\cite{changfano}.

Previous applications of the method outlined in Ref.~\cite{hamilton} have
been on indirect DR in which electron-impact dissociation is not energetically
open.  Thus, the outgoing-wave vibrational states have corresponded to closed
channels in these studies.  In contrast, 
for the present analysis of the high-energy peak in HeH$^+$ DR, 
electron-impact dissociation to H$^+$ + He
is energetically open (that to H + He$^+$ may be) 
and we have outgoing-wave vibrational functions as
open channels.  

It is worth mentioning that in either case, the asymptotic form
of the wavefunction is unphysical, because of the presence of outgoing
wave basis functions with complex energy that are exponentially 
increasing in magnitude. 
For the frame-transformed S-matrix in the c-normed vibrational basis, 
each column of which corresponds to an outgoing-wave scattering 
state $\Psi^{+(i)}$,
\begin{equation}
\Psi^+_\alpha = \chi_{\alpha}(R) h^-_{l_\alpha}(k_\alpha r)  + \sum_\beta S_{\alpha \beta} \chi_{\beta}(R) h^+_{l_\beta}(k_\beta r)  \ ,
\end{equation}
the proper statement of unitarity easily is found not to be
\begin{equation}
\forall_\alpha \quad \sum_\beta \big\vert S_{\alpha \beta} \big\vert^2 = 1
\end{equation}
but
\begin{equation}
\forall_\alpha \quad \sum_{\beta\gamma} S_{\alpha\beta} U_{\beta\gamma} S_{\alpha\gamma}^*  = 1
\end{equation}
where
\begin{equation}
\qquad U_{\beta\gamma} = \int \ dR \ \chi_\beta(R) \chi_\gamma(R)^* \ .
\end{equation}
While it is easy to prove that the frame-transformed and channel-closed 
S-matrix with bound
states is unitary in the complete basis limit, we have no proof that our
constructed S-matrix is guaranteed to be subunitary, though we have found
this to be the case in the applications of this theory to date.

Thus, the outline of the calculation is as follows.  We calculate
fixed-nuclei S-matrices $s_{lm, l'm'}(R)$ and transform them to the
ECS basis via Eq.(\ref{trans}).  We then
apply the rotational frame transformation
of Chang and Fano~\cite{changfano}.
 We finally apply the channel-closing formula,
\begin{equation}
\begin{split}
\mathscr{S}(E) & = S_{oo} - S_{oc}\left( S_{cc} - e^{-2i\beta(E)} \right)^{-1} S_{co} \\
\beta(E) & = \frac{\pi \delta_{\alpha \beta}}{\sqrt{2(E_\alpha-E)}} \ , \\
\end{split}
\label{closing}
\end{equation}
where the subscript $c$ and $o$ denote the closed and open channel subblocks of the
MQDT S-matrix  $S$, $\alpha$ is now a collective index including angular
momentum quantum numbers, and we introduce the notation $\mathscr{S}$ for the physical S-matrix.  The DR cross section is then obtained 
from the unitary defect of the frame-transformed S-matrix,
\begin{equation}
\sigma_\alpha(E) = \frac{\pi}{2E} \left( 1 - \sum_{\beta\gamma} \mathscr{S}_{\alpha\beta} U_{\beta\gamma} \mathscr{S}_{\alpha\gamma}^* \right) \ .
\end{equation}
We Boltzmann average and convolute with respect to the parallel and
transverse spreads in the incident electron energy in the experiment,
as described in Ref.~\cite{roman2}.

\begin{figure*}
\begin{center} 
\begin{tabular}{ccc}
\resizebox{0.52\columnwidth}{!}{\includegraphics*[0.9in,0.6in][5.3in,4.0in]{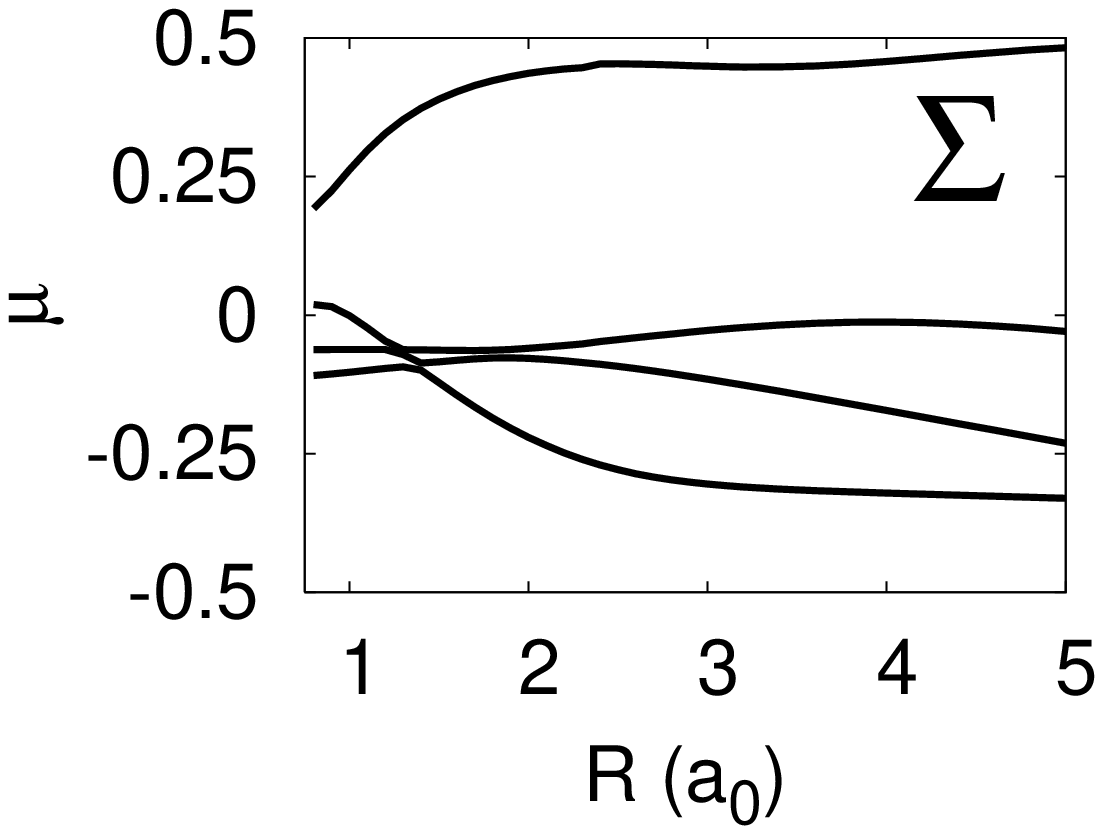}} &
\resizebox{0.52\columnwidth}{!}{\includegraphics*[0.9in,0.6in][5.3in,4.0in]{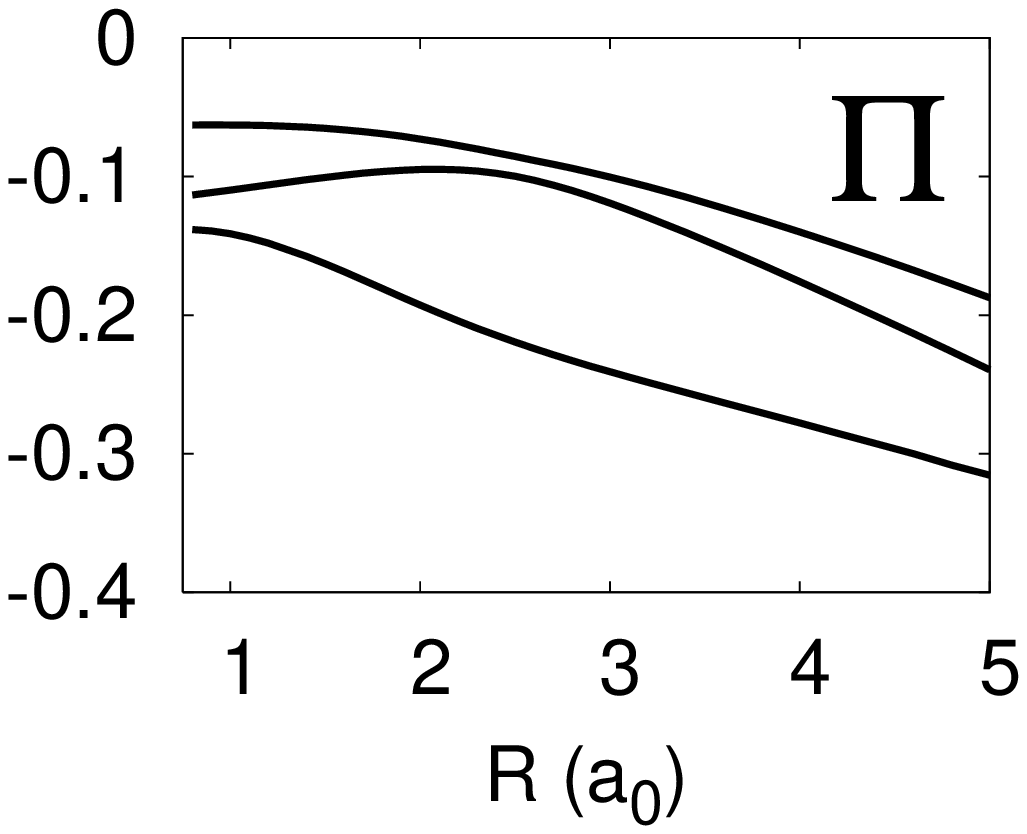}} &
\resizebox{0.52\columnwidth}{!}{\includegraphics*[0.9in,0.6in][5.3in,4.0in]{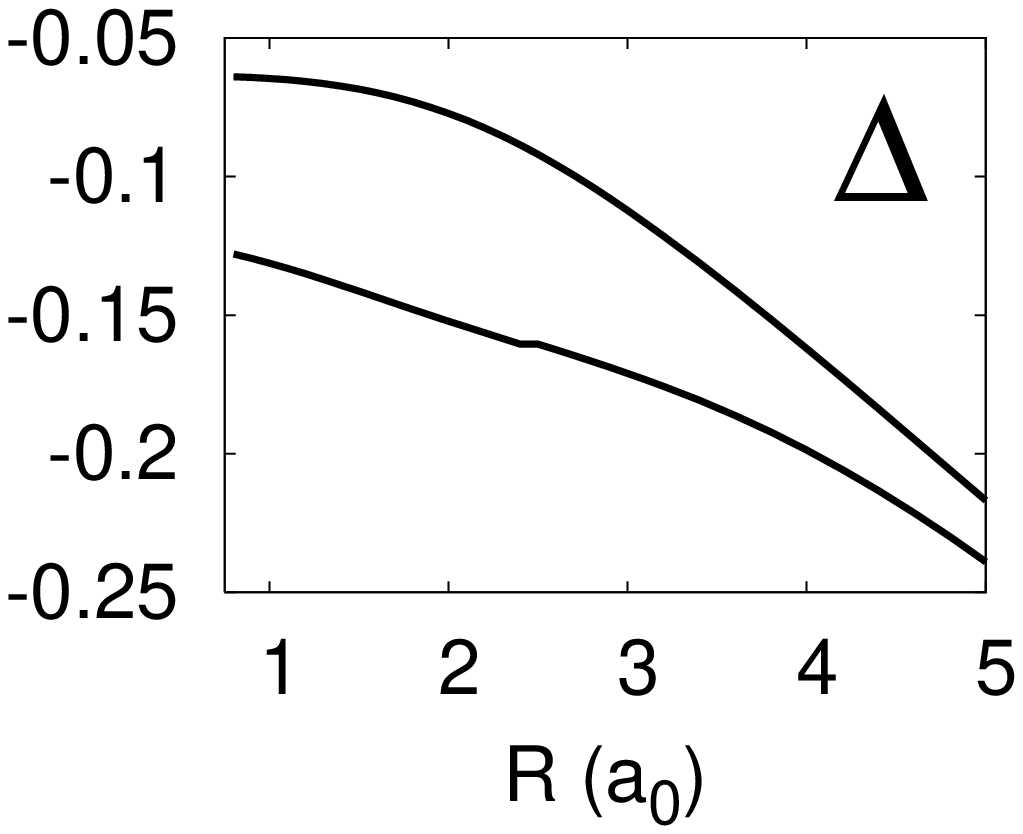}} 
\end{tabular}
\end{center}
\caption{Quantum eigendefects calculated at zero incident energy in the ground state cation channel.
\label{defects}}
\end{figure*}

\section{Calculation of \textit{ab initio} quantum defects and cation rovibrational states}

We calculate quantum defects using the polyatomic 
UK R-matrix program\cite{ukrmatrix}, which is based on the 
Swedish-Molecule and Alchemy suites of quantum chemistry codes\cite{alchemy}.
We include $s$, $p$, $d$, and $f$-wave basis functions (only $s$, $p$, and
$d$ for the high-energy peak) for the scattered 
electron in an R-matrix box of radius 14 bohr, calculated with the
program GTOBAS\cite{gtobas}.  We use the cc-pvtz basis set.
For calculations on 
the low-energy region, we include the lone 1$\sigma$ SCF orbital of 
HeH$^+$ and an additional six $\sigma$ and two pairs of $\pi$ virtual 
orbitals for the target space.  The three lowest cation target states,
1 and 2 $^1\Sigma$ and $^3\Sigma$,
are included in the scattering calculation and defined by full CI
in the target space.  Penetration terms defined by full CI of three
electrons in the target space are included in the scattering wavefunction.
   
For the high-energy DR calculations, the HeH$^+$ target states are 
represented by full CI in the
orbital space
of a CASSCF calculation on the ground state, performed with the molpro quantum 
chemistry package\cite{MOLPRO}.  Following Orel 
\textit{et al.}\cite{rescignoheh}, we use a total of four $\sigma$
orbitals and one $\pi$ orbital.
We include the first three cation states and use an R-matrix radius
of eight bohr.  Penetration terms with full CI of three electrons are
again included in the scattering wavefunction.

Quantum defects at zero incident energy in the ground-state cation channel
are plotted in Fig.~\ref{defects}.

To perform the frame-transformation calculation to obtain the low-energy
DR rate, we interpolate the calculated fixed-nuclei s-matrices over a range of
$R$=0.75 to 4.5 bohr, and calculate outgoing-wave ECS vibrational states
using the finite-element Gauss-Lobatto discrete variable
representation \cite{femdvr}.  We used 16-point quadrature and seven elements,
scaling the $R$ coordinate at an angle of $\frac{1}{8}\pi$ on the sixth
element (the calculation is almost fully converged at 10th order
quadrature).  We included 42 vibrational wavefunctions; the calculations
of Ref.~\cite{pavanello} found 12 bound states, and our vibrational
states include seven with real energies and ten with imaginary energy
component less than 0.0001 hartree in magnitude.
We use the ground-state cation potential energy surface of Coxon 
and Hajigeorgiou\cite{coxon} and use modified atomic weights with reduced mass
\begin{equation}
\mu_{eff} = \frac{m_{H^{\frac{1}{2}+}}  m_{He^{\frac{1}{2}+}}}{m_{H^{\frac{1}{2}+}} + \ m_{He^{\frac{1}{2}+}}} \ ,
\end{equation}
where $m_{He^{\frac{1}{2}+}}$ is the mass of a helium
nucleus plus one and a half electrons and $m_{H^{\frac{1}{2}+}}$ is
the mass of a proton plus a half electron.
 The results are slightly sensitive to such fine-tuning
(which approximately accounts for nonadiabatic coupling) and we find this choice
to produce the best agreement with prior calculations of HeH$^+$ rovibrational
spectra.
For transitions up to $\nu=3$ or $j=3$ we find
good agreement with the results of Bishop and Cheung\cite{bch}, with errors
less than 21 cm$^{-1}$.

\begin{figure}
\resizebox{0.92\columnwidth}{!}{\includegraphics*[0.7in,0.6in][5.85in,4.2in]{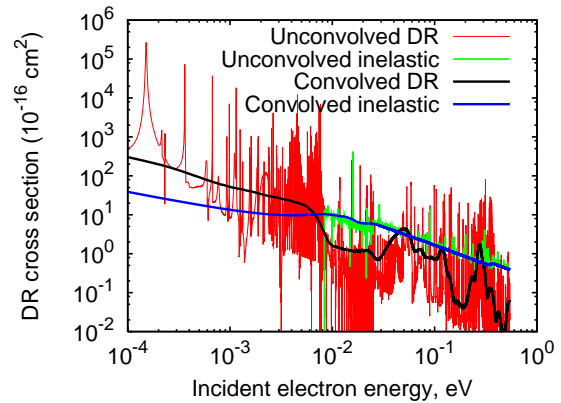}}
\caption{(Color online) Unconvolved cross section calculated for the low-energy DR of the ground rovibrational state of $^4$HeH.\label{unconv}}
\end{figure}

\begin{figure}
\resizebox{0.92\columnwidth}{!}{\includegraphics*[0.7in,0.6in][5.85in,4.2in]{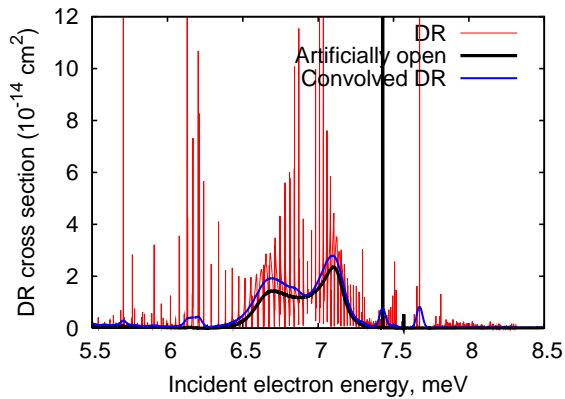}}
\caption{(Color online) Cross section calculated in the vicinity of the $j=1$ threshold at 8.3meV.  Unconvolved physical DR cross section, red line; physical DR cross section convolved with a gaussian, purple line; unphysical cross section calculated by artificially opening the $j=1$ channel, black line.\label{unclosed}}
\end{figure}

\begin{figure}
\begin{center}
\begin{tabular}{c}
\resizebox{0.90\columnwidth}{!}{\includegraphics*[0.7in,1.22in][5.85in,4.0in]{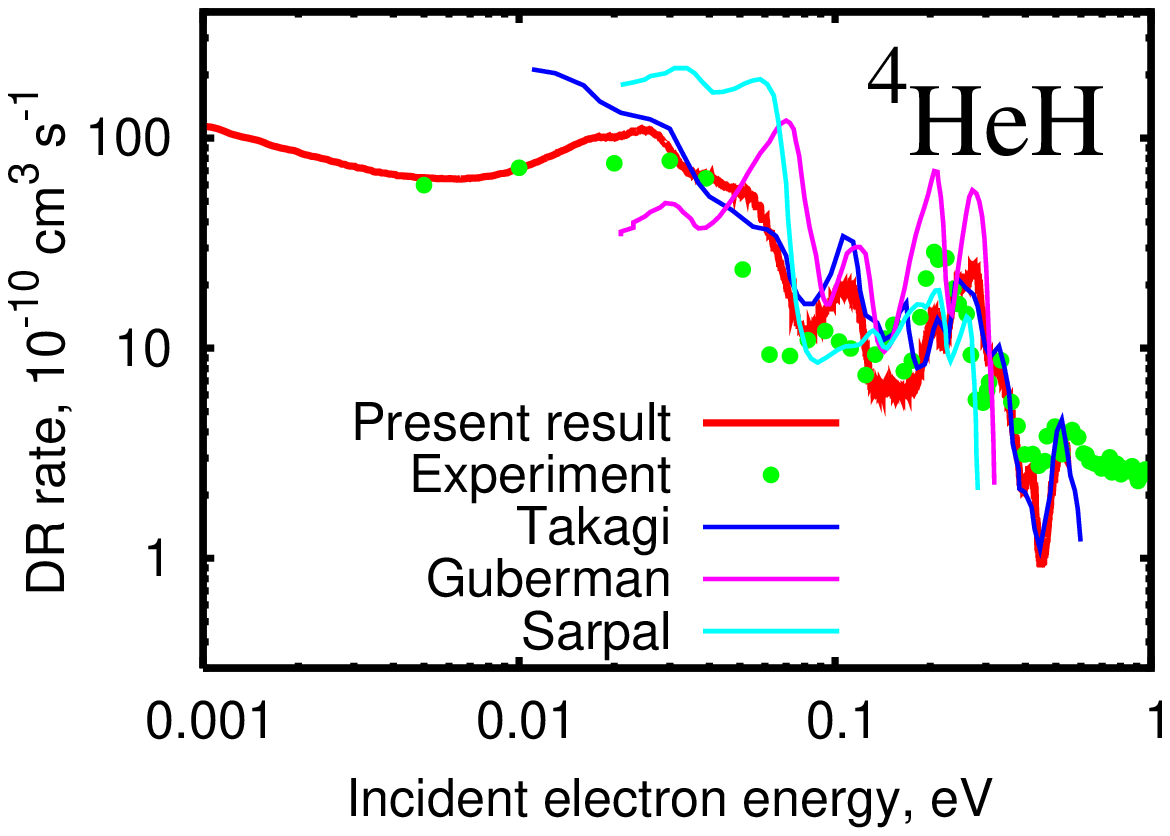}} \\
\resizebox{0.90\columnwidth}{!}{\includegraphics*[0.7in,1.22in][5.85in,4.0in]{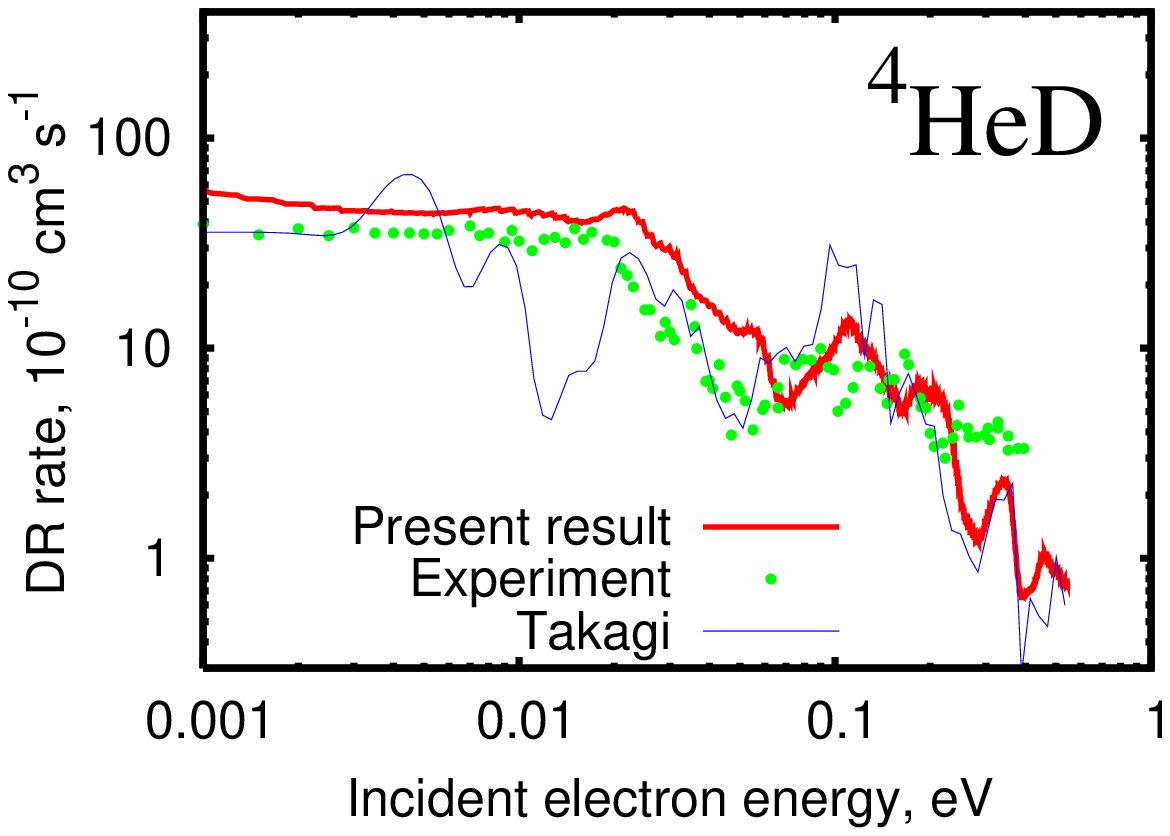}} \\
\resizebox{0.90\columnwidth}{!}{\includegraphics*[0.7in,1.22in][5.85in,4.0in]{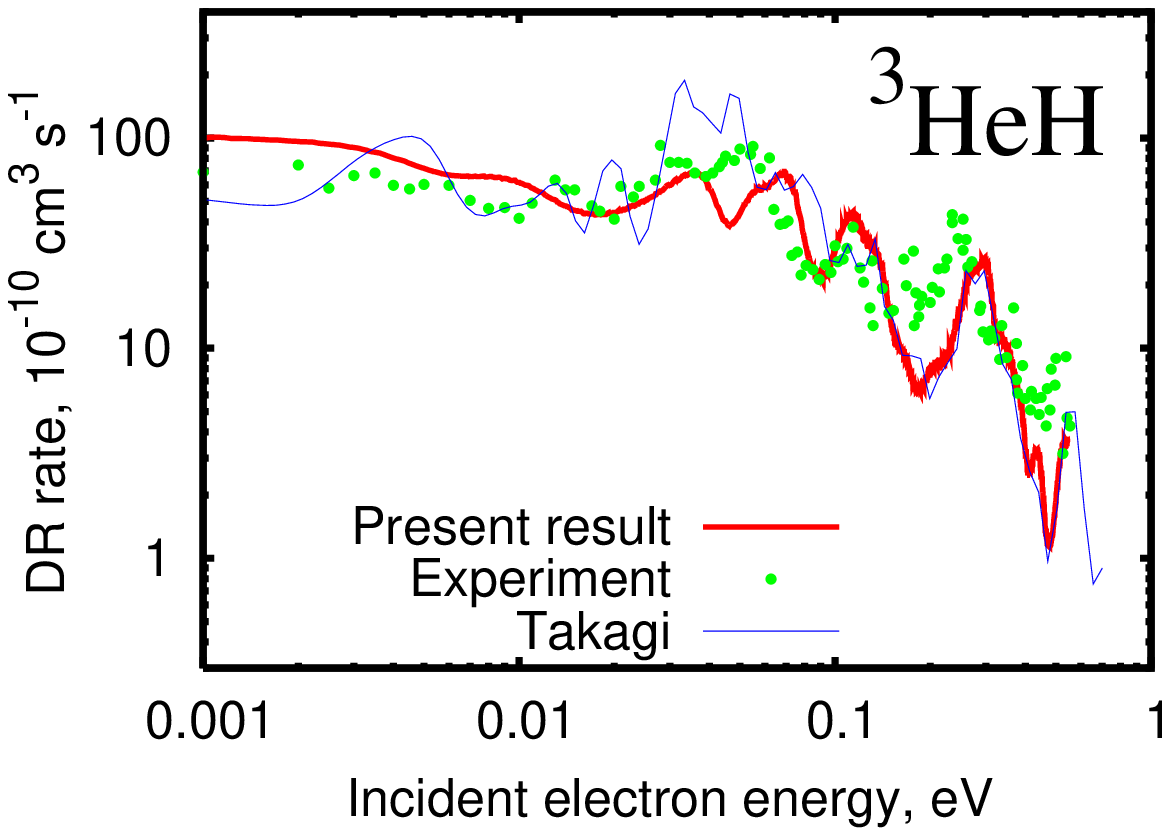}} \\
\resizebox{0.90\columnwidth}{!}{\includegraphics*[0.7in,0.7in][5.85in,4.0in]{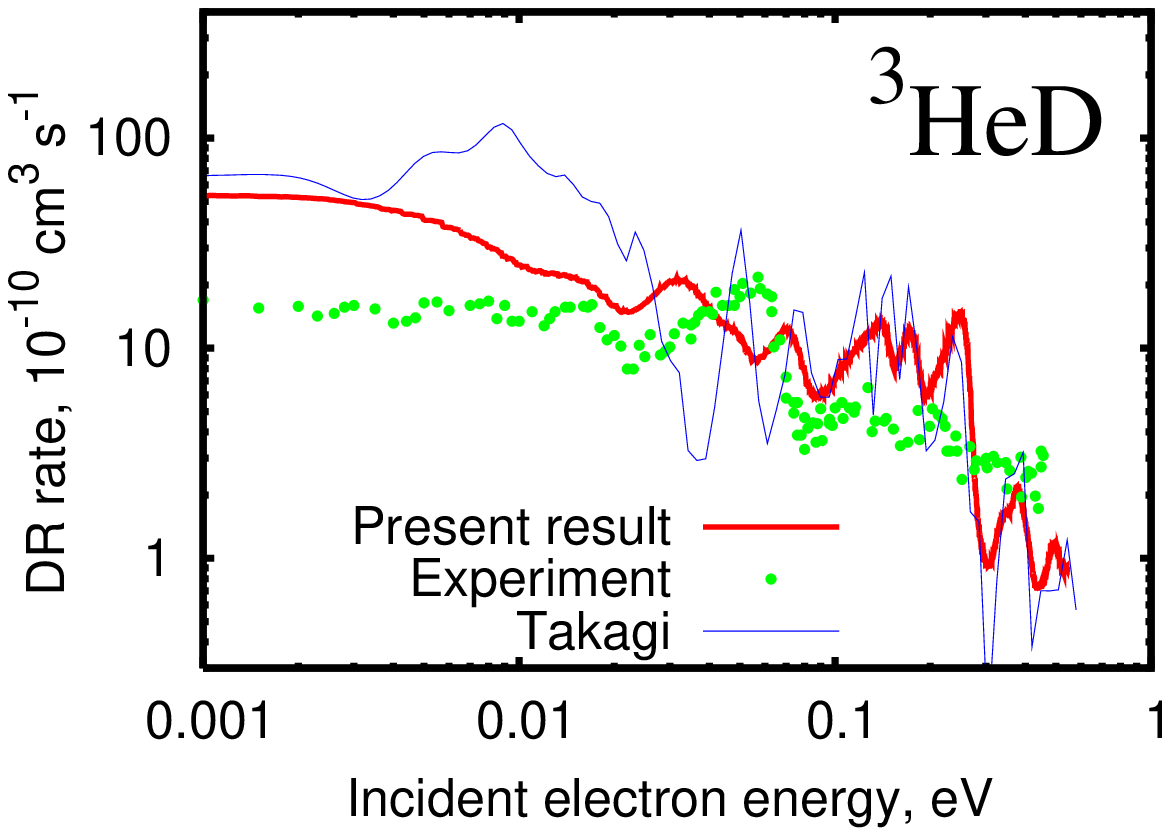}}
\end{tabular}
\end{center}
\caption{(Color online) Low-energy DR rate coefficient compared with the experiment of Tanabe \textit{et al.}\cite{tanabe} and the theoretical results of Takagi\cite{tanabe}, Guberman\cite{guberman}, and Sarpal \textit{et al.}\cite{sarpal}. 
\label{lowefig}}
\end{figure}

\section{Results}
 
Results for the low-energy calculation on HeH$^+$ are plotted in Figs.~\ref{unconv}, \ref{unclosed}, and \ref{lowefig}.  Figure \ref{unconv} shows the unconvolved cross
section for DR of the ground rovibrational state of $^4$HeH.  To compare
with experiment, we also convolve our results using a transverse spread of 10meV and a parallel spread of 0.1meV in the incident electron energy, 
to account for the experimental parameters of Ref.~\cite{tanabe}.

Inelastic scattering leads to rotational
or vibrational excitation of the cation and is allowed starting
at an energy of 8.3meV, where the first rotational threshold ($j=1$) lies.
As shown in Fig.~\ref{unconv}, at that threshold there is a clear 
step down in the DR cross section,
and the inelastic cross section assumes a value near the averaged
DR rate below threshold.  This finding confirms the basic picture
of the indirect mechanism, and parallels our findings on the DR
of LiH$^+$~\cite{roman,roman2}.  Below the  $j=1$ threshold,
the only inelastic process open is dissociative recombination;
above it, rotational excitation of the cation is allowed, and this
process competes with DR and in fact dominates it, decreasing the
DR cross section.

Thus, for energies below the  $j=1$ threshold of 8.3meV, 
the DR is seen to be driven by capture into Rydberg states supported
by the  $j=1$ rotational state of the cation.  These resonances
are visible in Fig.~\ref{unconv} and have a sharp cut-off at the
8.3meV threshold.  Another forest of Rydberg states terminates
at the second excited rotational threshold, $j=2$, at 24.9meV; at this
energy, there is a dip visible in the convolved DR cross section,
but a series of broad resonances then appears and increases the
cross section.  The inelastic cross section has only a barely
visible step up at the second excited rotational threshold, showing
that rotational excitation to $j=1$ is much stronger than that to $j=2$.

A close-up view of the calculated cross section for DR of the ground
rovibrational state of $^4$HeH is shown in Fig.~\ref{unclosed}.  In this
figure we also plot the result of calculations in which the first
excited rotational channel ($j=1$) is left open even below its threshold.
Such a calculation elminates the Rydberg series converging to the
$j=1$ threshold and reveals the presence of resonances attached 
to higher channel thresholds.  In this figure we plot the unconvolved
physical DR cross section; the physical DR cross section convolved with
a gaussian to show its average value; and the unphysical, 
artificially-opened $j=1$
channel result. 
Note that, as in Refs.\cite{roman,roman2},
there is an enhancement of the DR rate at a complex resonance,
i.e., a resonance involving a lower principal quantum number 
state embedded in the very 
high Rydberg series converging to an excited cation state threshold.  In 
contrast to Refs.\cite{roman, roman2}, however, 
the complex resonance mechanism is here seen to be supported by
rotationally autoionizing resonances, rather than vibrationally
autoionizing ones.
This suggests that some of the ideas shown in 
Refs.\cite{roman,roman2}, concerning the indirect 
mechanism of DR being dominated by complex resonances, may also be relevant 
for rotationally-autoionizing states in addition to the originally-postulated
situation of vibrationally-autoionizing Rydberg states.  It is beyond the 
scope of this study to examine this point in greater detail, but this point 
is worth exploring in future research.

The results presented above analyze the DR of rotationally and
vibrationally cold, ground-state $^4$HeH.  To compare
with current experiments, we must account for the nonzero temperature
of the ion source.  We Boltzmann-average over the population
of rovibrational states at 800$^\circ$ K,
to account for the experimental parameters of Ref.\cite{tanabe},
and show the results for the various isotopologues in
Fig.~\ref{lowefig}.
We compare with the theoretical results of Guberman\cite{guberman}, Sarpal \textit{et al.}\cite{sarpal}, and Takagi\cite{tanabe} and it seems that
the present method has produced the results most closely matching the
experimental data of Tanabe\cite{tanabe}.

For calculating the high-energy DR peak, several degrees of complication
may be included in the calculation, but we find that a very simple treatment
is sufficient to approximate the experimental results.  
We include neither the energy dependence nor the R-dependence 
of the fixed-nuclei S-matrix, evaluating
it at 0.6 hartree and 1.6$a_0$, roughly at the center of the experimentally observed peak and ground-state vibrational wavefunction.

\begin{figure}
\begin{center}
\begin{tabular}{c}
\resizebox{0.85\columnwidth}{!}{\includegraphics{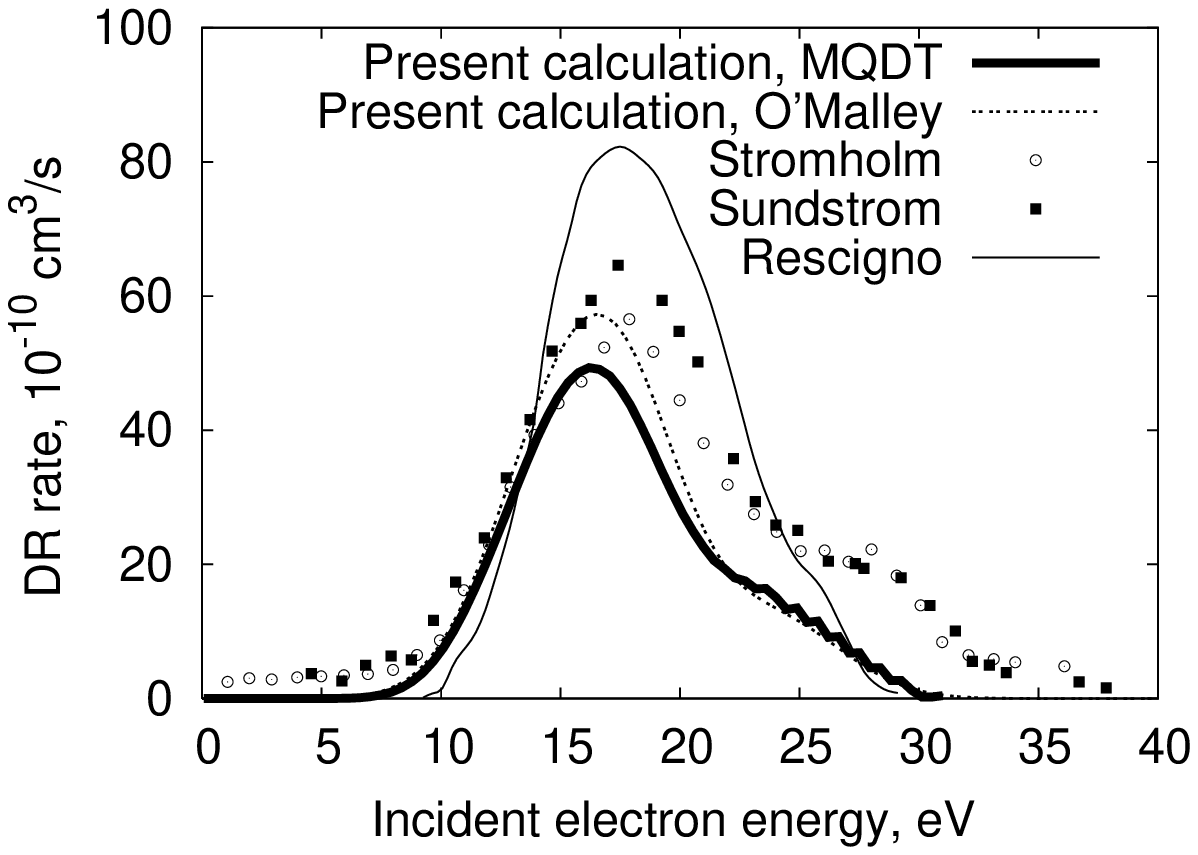}} \\
\resizebox{0.85\columnwidth}{!}{\includegraphics{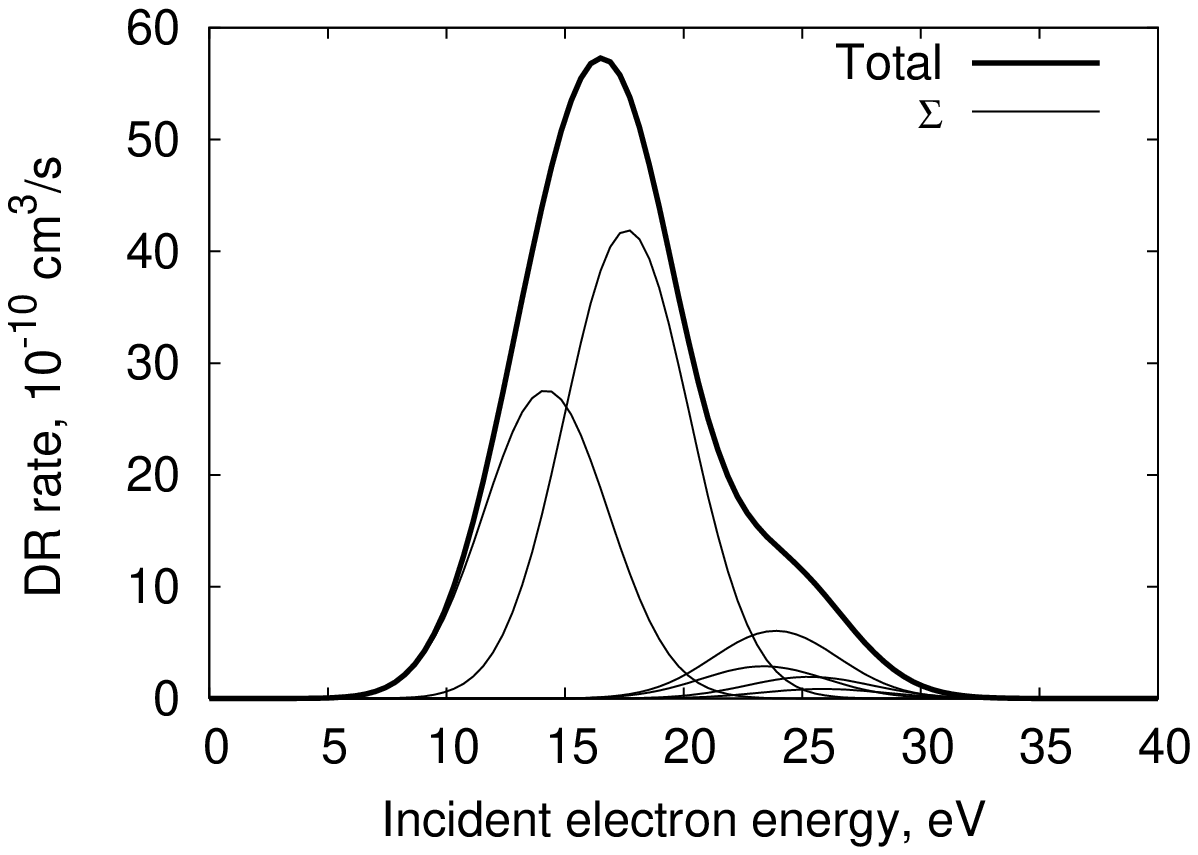}}
\end{tabular}
\end{center}
\caption{\textbf{Top:} High-energy dissociative recombination rate 
coefficient calculated with the
present MQDT treatment, compared with our results using calculated 
resonance curves and the O'Malley treatment (Eq.(\ref{omalleyeqn})), the theoretical results of
Orel \textit{et al.}\cite{rescignoheh}, and the experimental results of 
Sundstrom \textit{et al.}~\cite{sundstrom} and 
Stromholm \textit{et al.}~\cite{stromholm}.
\textbf{Bottom:} Contribution of each of the six sigma resonances to our
total result using the O'Malley treatment.
\label{highefig}}
\end{figure}

Results for the high-energy DR calculation are shown in Fig.~\ref{highefig}.  We compare
our MQDT results to the prior theoretical results of Ref.~\cite{rescignoheh}
and the experiments of Refs.~\cite{sundstrom} and \cite{stromholm}.  We
also calculate a result using the O'Malley framework~\cite{omalley} in which
we first find the resonance positions and widths as a function of
nuclear geometry.  We do so by applying the MQDT channel-closing formula,
Eq.(\ref{closing}), to the calculated MQDT S-matrices to obtain the
"physical" S-matrix $\mathscr{S}(E)$ in the complex-energy plane.
We then locate its poles $E_R^{(i)} - i\frac{\Gamma^{(i)}}{2}$.
We find six resonances, $i=1-6$, all of which have sigma symmetry.  Using the
approximate formula~\cite{omalley} valid for small widths,

\begin{equation}
\sigma(E) = \sum_i \frac{\pi^2}{E} \frac{\Gamma^{(i)}(R_i(E))}{\left\vert {E_R^{(i)}}'(R_i(E))\right\vert} \left\vert \Psi(R_i(E)) \right\vert^2 \ ,
\label{omalleyeqn}
\end{equation}
where the widths $\Gamma^{(i)}$, resonance energies $E_R^{(i)}$, and 
the ground-state vibrational wavefunction $\Psi$ are evaluated at a geometry
$R_i(E)$ consistent with a vertical transition such that $E_0 + E = E_R^{(i)}(R_i(E))$, with $E_0$ the initial state energy, 
we obtain a result consistent with our MQDT result.  In the lower
panel of Fig.~\ref{highefig} we show the
contributions for each of the six sigma resonances to the total.
 Orel
\textit{et al.}\cite{rescignoheh} found two $\Pi$ resonances having significant contribution
to the total cross section, and their six $\Sigma$ resonances give
partial cross sections different from ours shown in Figure~\ref{highefig}.

\section{Discussion}

We have applied the theory of Ref.~\cite{hamilton} to the calculation of
DR rates for HeH$^+$ + e$^-$ in both the low- and high-energy regions.
We have achieved very good agreement with experiment -- at least as good
as that of Orel \textit{et al.}~\cite{rescignoheh} 
for the high-energy peak, and better than all prior calculations for
the low-energy region.  Thus, it is clear that the present methodology gives
accurate results for both the indirect and direct mechanisms.

Our treatment of the high-energy peak was perhaps the simplest possible:
we used a quantum defect matrix constant with respect to both incident
electron energy and internuclear radius.  Such a simple treatment is 
applicable to much larger systems, as it leads to a sparse system of
linear equations in the channel closing step, which is the rate limiting
step for polyatomic DR calculations.

However, for larger systems it may be necessary to include the
energy dependence of the quantum defect in order to reproduce 
the relvant physics of DR or rovibrational autoionization.
For instance, larger polyatomic systems such as NO$_2^+$ will
in general support a greater number of shape resonances than
smaller diatomic systems, and these will impart energy-dependence
to the quantum defect functions that is difficult to analytially
remove.

Methods suited to a MQDT treatment including the energy dependence
of the fixed-nuclei quantum defect include those of
Refs.\cite{greenevib, gaogreene}.  These calculations require not 
the S-matrix but other entities such as the square root of the S-matrix
or the quantum defect matrix.  Therefore, such a calculation is more difficult
owing to the necessity of following the branches of the quantum defect
matrix across both $R$ and $E$, for a large ($>$ 1 hartree) energy range.

Even for the relatively small HeH$^+$ system,
the high-energy DR peak spans a large energy range and involves a large
transfer of energy from the electronic to the nuclear degrees of freedom.
Given these qualities,
it is not \textit{a priori} obvious that our simple treatment would
accurately reproduce the physical results.  We pursued an energy-dependent
treatment along the lines of~\cite{gaogreene}, and this was yielding
results similar to those presented in Fig.\ref{highefig}, but was
not numerically stable, and would require improvement to yield
publishable results.

\section{Acknowledgments}
 
We thank A. E. Orel for helpful suggestions and encouragement.  
This work was supported in part by the DOE Office of Science and by NSF 
grant number ITR 0427376, and one of us 
(CHG) received partial support from the Alexander von Humboldt Foundation.

\end{document}